\documentclass[aps,preprint,superscriptaddress,a4paper,floatfix]{revtex4-1}

\usepackage{amssymb}
\usepackage{graphicx}
\usepackage{amsmath}
\usepackage{latexsym}
\usepackage{xspace}
\usepackage{bm}

\newcommand{\kappamn}{$\kappa$-(BETS)$_2$\-Mn\-[N(CN)$_2$]$_3$\xspace}
\newcommand{\muh}{\mu_0 H}

\begin{document}

\title{Interplay between the $d$- and $\pi$-electron systems in magnetic torque of the layered
organic conductor $\kappa$-(BETS)$_2$Mn[N(CN)$_2$]$_3$}

\author{O.~M.~Vyaselev}
\email{vyasel@issp.ac.ru}
\affiliation{Institute of Solid State Physics, Russian Academy of Sciences,
Academician Ossipyan str. 2, 142432 Chernogolovka, Moscow region, Russia}

\author{W.~Biberacher}
\affiliation{Walther-Mei{\ss}ner-Institut, Bayerische Akademie der Wissenschaften,
Walther-Mei{\ss}ner-Str. 8, 85748 Garching, Germany}

\author{N.~D.~Kushch}
\affiliation{Institute of Problems of Chemical Physics, Russian Academy of Sciences,
Academician Semenov ave. 1, 142432 Chernogolovka, Moscow region, Russia}

\author{M.~V.~Kartsovnik}
\email{mark.kartsovnik@wmi.badw-muenchen.de}
\affiliation{Walther-Mei{\ss}ner-Institut, Bayerische Akademie der Wissenschaften,
Walther-Mei{\ss}ner-Str. 8, 85748 Garching, Germany}

\date{\today}

\begin{abstract}

In the organic charge transfer salt \kappamn the metallic conductivity is provided by
itinerant $\pi$-electrons in the layers of BETS molecules, whereas magnetization is
largely dominated by the localized $d$-electrons of the Mn$^{2+}$ ions in the insulating
anionic layers. We study magnetic properties of the compound in its low-temperature,
Mott-insulating state by means of magnetic torque technique. The complex behavior of the
torque can be qualitatively explained by the coexistence of two weakly interacting
magnetic subsystems associated with paramagnetic $d$-electron spins and antiferromagnetically
ordered $\pi$-electron spins, respectively. Based on the experimental data, we determine
the principal axes of magnetization of the Mn$^{2+}$ sublattice and propose a qualitative
model for the $\pi$-electron spin arrangement, implying an important role of the
Dzyaloshinskii-Moriya interaction.

\end{abstract}

\maketitle

\section{Introduction}
The organic radical cation salt  $\kappa$-(BETS)$_2$Mn[N(CN)$_2$]$_3$, where BETS stands for
bis-(ethylenedithio)tetraselenafulvalene, has a layered structure
consisting of conducting sheets of BETS donor molecules, sandwiched between
insulating Mn[N(CN)$_2$]$_3^-$ anion layers \cite{KJACS08,ZvPR10}.
This compound adds to the series of BETS salts with spatially separated conducting and magnetic
systems synthesized in a quest for hybrid multi-functional molecular materials combining conducting
and magnetic properties in the same crystal lattice, potentially promising for microelectronics.
The earlier members of this family, $\lambda$- and $\kappa$-(BETS)$_2$FeX$_4$ (X=Cl, Br)
\cite{koba96,akut98,FuJACS01,KobCR04} have been of strong interest
due to prominent interactions between the localized $d$-electron spins of
the Fe$^{3+}$ ions in the insulating layers and itinerant $\pi$-electrons in the conducting BETS
layers. For example, a considerable $\pi$--$d$ coupling in $\lambda$-(BETS)$_2$FeCl$_4$ leads
to a metal-insulator transition in the $\pi$-electron system triggered by an antiferromagnetic (AF)
ordering of localized Fe$^{3+}$ spins \cite{BrossEPJ98} and to a spectacular phenomenon of
superconductivity induced by a strong magnetic field \cite{uji01a,bali01}. In the $\kappa$-(BETS)$_2$FeX$_4$
salts the $\pi$--$d$ coupling is weaker; however it can be readily traced in a
reconstruction of the Fermi surface caused in the AF state
\cite{kono05,kono07,kunz16}, high-field re-entrant superconductivity \cite{FuJACS02,kono04b},
and protection of the low-field superconductivity by the AF ordering \cite{kono04b,kart16}.

In the present compound the $\pi$--$d$ interactions seem to be even weaker. While the
metal-insulator transition at $T_{\mathrm{MI}}\approx 21$\,K \cite{KJACS08} might, at first glance,
appear similar to that in $\lambda$-(BETS)$_2$FeCl$_4$, it is most likely driven by purely the
Mott-insulating instability of the $\pi$-electron system and not by an AF instability of the
localized $d$-electron spins. Indeed, clear indications of a long-range AF ordering of the
itinerant $\pi$-electron spins have been obtained in NMR experiments at $T < T_{\mathrm{MI}}$
\cite{VyJL12, VyCr12}, whereas no sign for a long-range order was found for the Mn$^{2+}$ subsystem
\cite{VyPRB11, VyJETP11}.
Several anomalies associated with the metal-insulator transition have also been found
in magnetic torque experiments \cite{KJACS08, VyPRB11}. However, their exact origin has been
not clarified yet. Here we present a comprehensive study of magnetic torque in the insulating
state of $\kappa$-(BETS)$_2$Mn[N(CN)$_2$]$_3$ and discuss its behavior taking into account
the coexistence of the two weakly interacting spin  subsystems.

\section{Experimental}

The crystal structure of  \kappamn is monoclinic; the space group is $P$2$_1$/$c$ and the
lattice constants at 15\,K are: \textit{a}\,=\,19.421\,\AA, \textit{b}\,=\,8.346\,\AA,
\textit{c}\,=\,11.830\,\AA, $\beta$=92.90$^\circ$, \textit{V}=1915.0\,\AA$^3$, and $\rho=
2.424$\,g/cm$^3$, with two formula units per unit cell \cite{ZvPR10}. The conducting layers are
formed by BETS dimers in the $bc$-plane and sandwiched between the polymeric Mn[N(CN)$_2$]$_3^-$
anion layers in the \textit{a} direction. The crystal growth procedure and details of the structure
have been reported elsewhere \cite{KJACS08, ZvPR10}. Results of the magnetization measurements have
been reported previously \cite{VyPRB11}.

The sample was a 40\,$\mu$g thin-plate single crystal of $\sim 0.7\times 0.3\times 0.08\,$mm$^3$
size, with the largest dimensions along the conducting BETS layers (crystallographic $bc$-plane).
Magnetic torque was measured in fields up to 15\,T with a homemade cantilever beam
torquemeter described in \cite{ChrSSC94}. The cantilever was made of 50\,$\mu$m thick
as-rolled beryllium-copper foil. The torque was determined from the change of the capacitance
between the cantilever disc, to which the sample is attached, and the ground plate. The capacitance
was measured using a tunable capacitance bridge. The maximum torque of the cantilever produced by
the gravity force (in zero applied field) was $1.16\times 10^{-7}$\,N$\cdot$m, this value was used to convert the
measured changes in capacitance to the units of torque. The torquemeter was attached to a rotation stage
whose rotation axis was perpendicular to both the external magnetic field and the working plane
of the cantilever. In this geometry, the component of the torque along the rotation axis
is measured.

\section{Results}

\begin{figure}[h!]
\includegraphics[width=0.5\linewidth]{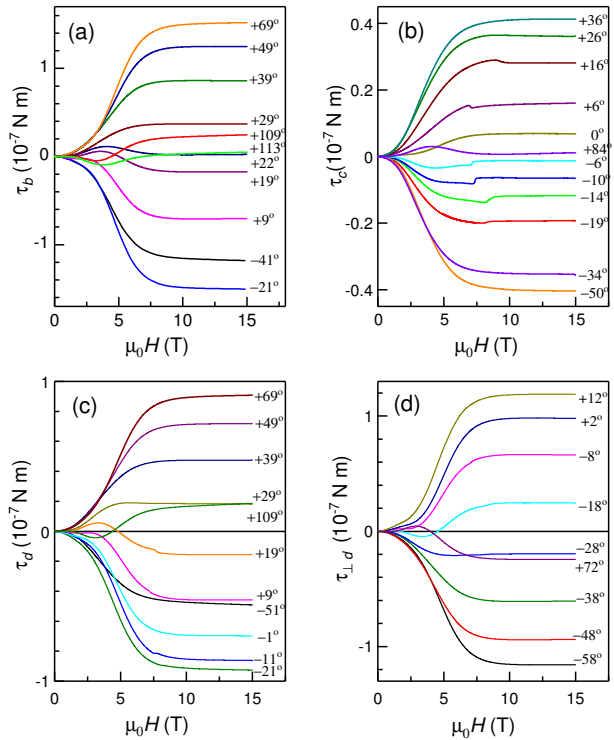}
\caption{(Color online) Field dependence of the magnetic torque of  \kappamn measured at
$T=1.5$\,K for the rotation axis parallel to directions: (a) [$0\bar{1}0$], (b) [001],
(c) [$0\bar{1}1$], and  (d) perpendicular to [$0\bar{1}1$] in the $bc$-plane. Numbers to the right
of the curves indicate the polar angle $\theta$ between the field direction and $a^\ast$, the normal
to the crystallographic $bc$-plane. \label{HDeps}}
\end{figure}

Panels (a)\,--\,(d) of Fig.\,\ref{HDeps} show the magnetic field dependence of the
torque, $\tau(H)$, on the \kappamn crystal measured at 1.5\,K, with the rotation axis parallel,
respectively, to [$0\bar{1}0$] ($\tau_b$), [001] ($\tau_c$), [$0\bar{1}1$] ($\tau_d$), and
perpendicular to [$0\bar{1}1$] in the $bc$-plane ($\tau_{\perp d}$). Numbers to the right
of the curves indicate the polar angle $\theta$ between the field direction and $a^\ast$,
the direction perpendicular to the crystallographic $(bc)$ plane.

There are several notable features in Fig.~\ref{HDeps}, which will be discussed below:

\noindent (i) At high fields ($\muh >10$\,T) the torque becomes constant in field;

\noindent (ii) For the angles where the high-field torque is small, see, e.g. the $\theta = 22^{\circ}$
curve for $\tau_b$ in Fig.~\ref{HDeps}(a) or  the $\theta = -6^{\circ}$ curve for $\tau_c$ in
Fig.~\ref{HDeps}(b), the torque is nonmonotonic in the range between $\simeq 2.5$ and 7.5\,T;

\begin{figure}[h!]
\includegraphics[width=0.5\linewidth]{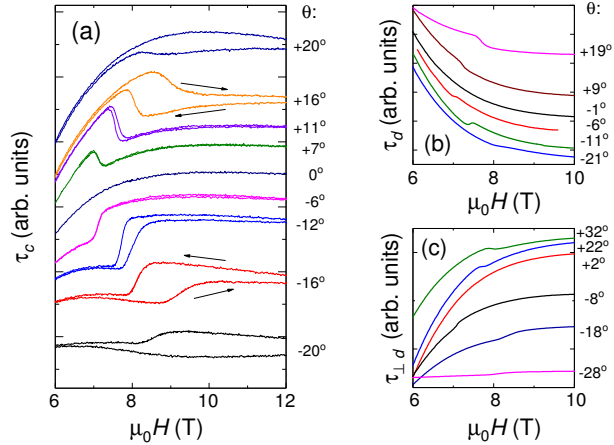}
\caption{(Color online) A close-up of the steplike features (kinks) in the $H$-dependence of $\tau_c$
(a), $\tau_d$ (b) and  $\tau_{\perp d}$ (c). The curves are shifted along the vertical axis for
clarity. \label{Kinks}}
\end{figure}

\noindent (iii) At some angles $\tau_c$, $\tau_d$ and $\tau_{\perp d}$
demonstrate a steplike feature (``kink'') at fields 7-10\,T. Figure~\ref{Kinks}
shows the kinks in more details. No such kinks have been detected for $\tau_b$ at any
$\theta$.

Features (ii) and (iii) vanish as the temperature is increased above $T_\mathrm {MI}$:
the kinks disappear \cite{VyPRB11}, the field dependence becomes monotonic and gradually
acquires the simple parabolic form usual for an
anisotropic paramagnet at $\mu_B B\ll k_B T$ (where $\mu_B$ is the Bohr magneton and $k_B$ is the
Boltzmann constant). Therefore, these features must be associated with the low-temperature
insulating state with antiferromagnetically ordered $\pi$-electron spins.

On the other hand, the field directions where the high-field torque is zero $\muh=15$\,T,
e.g. $\theta\simeq 22^{\circ}\pm 90^{\circ}$ for $\tau_b$, Fig.~\ref{HDeps}(a), or
$\theta\simeq -6^{\circ}\pm 90^{\circ}$ for $\tau_c$, Fig.~\ref{HDeps}(b), are at
$T=1.5$\,K, the same as at high temperatures ($T>T_\mathrm {MI}$) within the
experimental accuracy $\pm 0.5^{\circ}$. This means that the principal axes of the
high-field magnetization above and below $T_\mathrm {MI}$ coincide.

\section{Discussion}

The absolute values of torque in Fig.\,\ref{HDeps} are more than an order of magnitude higher
than in the structurally similar but free of magnetic ions charge-transfer salt
$\kappa$-(BEDT-TTF)$_2$Cu[N(CN)$_2$]Cl \cite{pint99}.
In turn, the kinks have been related to the antiferromagnetically ordered $\pi$-electron spins
\cite{VyPRB11}. In what follows we characterize the phenomena associated with each spin subsystem
separately and address implications of their interaction.

\subsection{General expressions for the magnetic torque.}

The magnetic torque is expressed as
\begin{equation}
\bm{\tau}=V\bm{M}\times\bm{B}\,, \label{tau}
\end{equation}
where $V$ is the volume of the sample, $\bm{M}$ is the sample magnetization and $\bm{B} =
\mu_0\bm{H}+ \mu_0 \bm{M}$ is the magnetic field. Let us neglect for a while the ramifications due
to the sample shape (that will be discussed below) and assume the sample is a sphere. In that case
\begin{equation}
\bm{\tau}=V\mu_0\bm{M}\times(\bm{H}+ \bm{M})=V\mu_0\bm{M}\times\bm{H}. \label{tau1}
\end{equation}

Consider first the high-temperature, low-field limit, $\mu_B B\ll k_B T$. Assuming the field in the
$(XY)$ plane where $X$ and $Y$ are the magnetization principal axes,
\begin{equation}
\bm{H}=H[\cos\theta, \sin\theta,0]\,, \label{H}
\end{equation}
and the susceptibility tensor

\begin{equation}
\hat{\chi}=\left( \begin{array}{ccc}
\chi_X & 0 & 0 \\
0 & \chi_Y & 0 \\
0 & 0 & \chi_Z \end{array} \right), \label{chi}
\end{equation}

\noindent one obtains the magnetization:
\begin{equation}
\bm{M}=\hat{\chi} \cdot \bm{H} = H[\chi_X\cos\theta, \chi_Y \sin\theta, 0]\,, \label{M}
\end{equation}
and the torque  $\bm{\tau} = V \bm{M}\times \bm{H} = [0, 0, \tau_Z ]$, where
\begin{equation}
\tau_Z =\frac{1}{2} V H^2 (\chi_X-\chi_Y) \sin 2\theta\,, \label{lftau}
\end{equation}
which gives a quadratic in $H$ behavior of the torque at low fields/high temperatures,
consistent with the experiment at $\muh<2$\,T, see Fig.~\ref{HDeps}.

In the high-field, low-temperature regime, $\mu_B H\gg k_B T$, the linear field dependence
given by Eq.\,(\ref{M}) is no more valid. The magnetization of a paramagnet saturates,
and in a system with an isotropic $g$-factor the effect of changing $H$ reduces to
a change of the angle between the magnetization vector and the field direction.
In that case the axial anisotropy follows a
$H^{-2}$ law \cite{CorCCR01}, so that at $H\rightarrow \infty$  the torque asymptotically
approaches a constant value modulated by a $\sin 2\theta$ angular dependence. This behavior of the
torque is indeed observed in our experiment, as is seen in Fig.~\ref{HDeps} for $\muh>10$\,T.

However, the nonmonotonic field dependence of torque observed in the range 2.5-7.5\,T
and the kink features cannot be described within the model of an
anisotropic paramagnet but arise apparently due to the AF-ordered spins of
the $\pi$-electron subsystem, as discussed below.

\subsection{Principal axes of magnetization.}

We now proceed to determining directions of the principal axes of the magnetization in
\kappamn.

\begin{figure}[h!]
\includegraphics[width=0.5\linewidth]{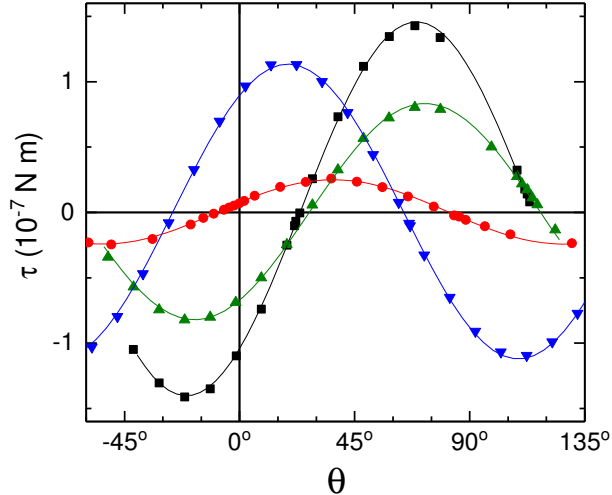}
\caption{(Color online) Angular dependence of the torque at 1.5\,K, 15\,T for the field rotated
around  $[0\bar{1}0]$ (squares), [001] (circles), $[0\bar{1}1]$ (up-triangles) and perpendicular
to $[0\bar{1}1]$ in the $bc$-plane (down-triangles).
Solid lines: fits to the data using Eq.\,\ref{exptau}.
\label{Angular}}
\end{figure}
Figure~\ref{Angular} shows the angle-dependent torque for different rotations at
$T=1.5$\,K, $\muh=15$\,T. The raw experimental data have been corrected for the demagnetization
effect as explained in the Appendix.

As one can see in Fig.~\ref{Angular}, all four curves follow nicely the dependence:
$\tau = \tau_{\mathrm{max}}\sin 2(\theta - \theta_0)$ with the paramters $\tau_{\mathrm{max}}$ and $\theta_0$
listed in Table~\ref{TableTau}.
For the practical reasons which will become clear below, it is more convenient to present this dependence in the
form:
\begin{equation}
\tau=\alpha \cos 2\theta+\beta \sin 2\theta\,,
\label{exptau}
\end{equation}
where $\alpha=-\tau_{\mathrm{max}}\sin 2\theta_0$ and $\beta=\tau_{\mathrm{max}}\cos 2\theta_0$.

\begin{table}[b]
\caption{\label{TableTau} Fit parameters to the torque data in Fig.~\ref{Angular}.}

\begin{ruledtabular}
\begin{tabular}{lccccc}
Rotation axis&$\phi$&$\tau_{\mathrm{max}}$[$10^{-7}$\,N$\cdot$m]&$\theta_0$&$\alpha$[$10^{-7}$\,N$\cdot$m] &$\beta$[$10^{-7}$\,N$\cdot$m]\\
\hline
$[0\bar{1}0]$ &0 & 1.43 & $24^\circ$&$-1.07$&0.948\\
$[001]$ & $90^\circ$& 0.246&$-8.4^\circ$&0.071&0.236\\
$[0\bar{1}1]$& $55^\circ$& 0.827&$27.5^\circ$&$-0.678$&0.474\\
$\perp[0\bar{1}1]$& $145^\circ$& 1.128&$-26^\circ$&0.888&0.696 \\
\end{tabular}
\end{ruledtabular}
\end{table}

In order to analyze the experimental results, we introduce the coordinate system $\{x,y,z\}$, where
$x$ is parallel to $a^{\ast}$ while $y$ and $z$ coincide with crystallographic $b$- and $c$-axes,
respectively. The rotation axis vector is given by $\bm{R} = [0,  -\cos \phi, \sin \phi]$, where
$\phi$ is the angle between the rotation axis and the $-\bm{b}$ direction. The values of $\phi$ for
the four reported rotations are listed in Table~\ref{TableTau}.

As mentioned above, at high field the linearity between $\bm{M}$ and $\bm{H}$ in the form of
Eq.\,(\ref{M}) is no more valid. In order to calculate the magnetization direction in this case,
instead of the susceptibility tensor $\hat{\chi}$ we introduce tensor $\hat{\xi}$  of the
directional cosines between $\bm{M}$ and $\bm{H}$ vectors,

\begin{equation}
\hat{\xi}=\left( \begin{array}{ccc}
d_{xy}+d_{xz}+d_{yz} & \xi_{xy} & \xi_{xz} \\
\xi_{xy} & -(d_{xy}-d_{xz})+d_{yz} & \xi_{yz} \\
\xi_{xz} & \xi_{yz} & d_{xy}-d_{xz}+d_{yz} \end{array} \right), \label{xi}
\end{equation}

\noindent where $d_{xy}=\xi_{xx} - \xi_{yy}$, $d_{xz}=\xi_{xx} - \xi_{zz}$, $d_{yz}=\xi_{yy} +
\xi_{zz}$. In that case $\bm{M}=M\hat{\xi}\cdot\bm{h}$, where $\bm{h}=[\cos \theta, \sin \theta
\sin \phi,\sin \theta \cos \phi ]$ is the applied field unit vector. In the $H\rightarrow \infty$
limit $\bm{M}$ aligns with $\bm{H}$, so that $(\hat{\xi}\cdot\bm{h})\cdot\bm{h}=1$. Then, since the
torque at high field is known to have a $K\sin 2\theta$ dependence where $K$ is a constant
\cite{CorCCR01}, instead of using by Eq.\,(\ref{tau1}) we express the torque as
\begin{equation}
\bm{\tau}=K/M (\bm{M}\times \bm{h}) = K(\hat{\xi}\cdot \bm{h})\times \bm{h}\,. \label{hftau}
\end{equation}

The torque component along the rotation axis, which is measured in the experiment,
is
\begin{equation}
\tau_{\phi}=\bm{\tau}\cdot\bm{R}=K \{-\cos 2\theta [\xi_{xz}\cos\phi + \xi_{xy}\sin\phi] + \sin
2\theta[d_{xy}+d_{xz}-(d_{xy}-d_{xz})\cos 2\phi - \xi_{yz}\sin 2\phi]/2 \}\,. \label{tauphi}
\end{equation}
For the four rotation axes used in the experiment we obtain:
\begin{subequations}
\label{tau4r}
\begin{equation}
\tau_b(\phi=0)=K \{-\xi_{xz}\cos 2\theta +d_{xz}\sin 2\theta \}\,, \label{tau0}
\end{equation}
\begin{equation}
\tau_c(\phi=90^{\circ})=K \{-\xi_{xy}\cos 2\theta +d_{xy}\sin 2\theta \}\,, \label{tau90}
\end{equation}
\begin{equation}
\tau_d(\phi=55^{\circ})=K \{-(0.82\xi_{xy}+0.57\xi_{xz})\cos 2\theta
+(0.67d_{xy}+0.33d_{xz}-0.47\xi_{yz})\sin 2\theta \}\,, \label{tau55}
\end{equation}
\begin{equation}
\tau_{\perp d}(\phi=145^{\circ})=K \{-(-0.82\xi_{xz}+0.57\xi_{xy})\cos 2\theta
+(0.33d_{xy}+0.67d_{xz}+0.47\xi_{yz})\sin 2\theta \}\,. \label{tau145}
\end{equation}
\end{subequations}

In fact, a detailed inspection of the sample orientation for the $c$-axis rotation has
revealed that the real direction of $c$-axis was slightly (by $\sim 4^{\circ}$) tilted
from the direction of the rotation axis, and the correct value for $\phi$  was $94^{\circ}$.
Taking this into account, we obtain the corrected value for $\tau_c$:
\begin{equation}
\tau_c(\phi=94^{\circ})=K \{-(0.998\xi_{xy}-0.07\xi_{xz})\cos 2\theta
+(0.995d_{xy}+0.005d_{xz}+0.07\xi_{yz})\sin 2\theta \}\,.
\tag{\ref{tau4r}e}
\end{equation}

Equating the fit parameters $\alpha$ and $\beta$ listed in Table~\ref{TableTau} to the
corresponding coefficients of $\cos 2\theta$ and $\sin 2\theta$  in Eq.\,(\ref{tau4r}), one obtains
the matrix:

\begin{equation}
K\hat{\xi}=\left( \begin{array}{ccc}
1.184+Kd_{yz} & 0 & 1.07 \\
0 & 0.712+Kd_{yz} & 0 \\
1.07 & 0 & -0.712+Kd_{yz} \end{array} \right). \label{Kxi}
\end{equation}

The magnetization principal axes are the eigenvectors of this matrix: $X=[\cos
\theta_M, 0, \sin \theta_M]$; $Y$=[0, 1, 0], $Z=[ -\sin \theta_M, 0,\cos \theta_M]$ with $\theta_M
= 24.2^{\circ}$ for any arbitrary $d_{yz}$. The $xz$-plane of the magnetization principal axes
coincides with the $ac$-plane of the crystal, which is quite reasonable since it is the mirror
plane of the crystal structure. The $X$ vector is directed at $24^{\circ}$ from the $a^{\ast}$
direction in the $ac$-plane.

As it was mentioned above, at high temperatures the directions of the field where the torque vanishes,
are the same as at $T=1.5$\,K, $\muh=15$\,T (Fig.~\ref{Angular}). This implies that the obtained
orientations of the principal axes of the magnetization are inherent to the Mn$^{2+}$ spin system
and do not change at the metal-insulator transition.

\subsection{Angular and Field dependence of the kinks.}

\begin{figure}[h!]
\includegraphics[width=0.5\linewidth]{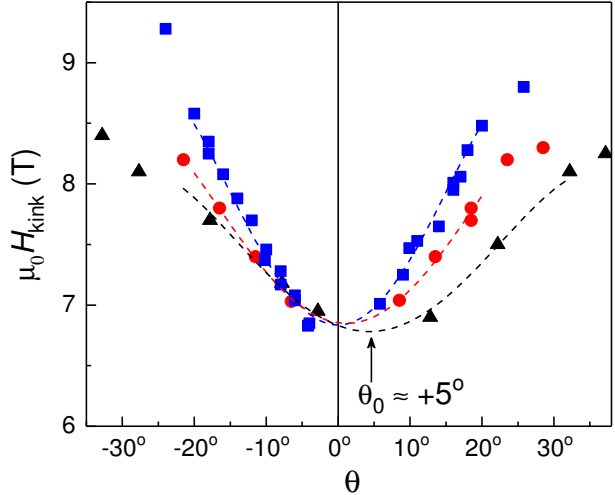}
\caption{(Color online) Angle dependence of the position of the kink feature in the field-dependent
torque $\tau_c$ (squares),  $\tau_d$ (circles) and  $\tau_{\perp d}$ (triangles).
\label{Hkink}}
\end{figure}

As one can see in Figs.\,\ref{HDeps} and \ref{Kinks}, the kink feature in the torque exists
when the field is tilted at a moderate angle, $|\theta | \lesssim 30^{\circ}$, from the
$a^{\ast}$ direction around the axis parallel to crystallographic directions $[001]$ or
$[0\bar{1}1 ]$ or to perpendicular to $[0\bar{1}1 ]$, but not around the $b$-axis ($[0\bar{1}0 ]$).
Figure~\ref{Hkink} shows the dependence of the kink position $H_{\mathrm{kink}}$ on the polar angle $\theta$
for the three above-mentioned rotation axes.

Thus, the following conditions should be satisfied in order to observe the kink:

\noindent $\bullet$ there must be a sufficiently large field component along $a^{\ast}$;

\noindent $\bullet$ there must be a component of the field along [010] (the b-axis);

\noindent $\bullet$ as mentioned above, the temperature must be below $T_\mathrm {MI}$.

\begin{figure}[b!]
\includegraphics[width=0.4\linewidth]{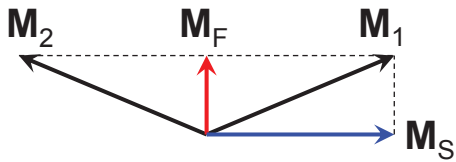}
\caption{(Color online) Presentation of the $(\bm{M}_1, \bm{M}_2)$ sublattice moments in the basis
of ferromagnetic $\bm{M}_F$ and staggered $\bm{M}_S$ magnetization vectors. \label{MsMf}}
\end{figure}

A very detailed description of the spin arrangement and field-induced spin reorientation (SR)
transition in another Mott-insulating organic salt, $\kappa$-(BEDT-TTF)$_2$Cu[N(CN)$_2$]Cl, which
has a structure similar to the present compound and undergoes an AF transition below $T_N = 27$\,K,
has been given in \cite{SmPRB03,SmPRL04}. The key concept is that in an AF system with a low
symmetry of the underlying crystal structure, the two magnetic sublattices $\bm{M}_1$ and
$\bm{M}_2$ do not arrange strictly antiparallel along the easy axis but form a \textit{canted}
antiferromagnetic (CAF) order due to the Dzyaloshinskii-Moriya (DM) interaction \cite{DzJETP57,
MorPR60}. Following the notations of Ref.\,\citenum{SmPRL04}, we introduce the ferromagnetic and
staggered magnetization vectors, which are expressed through the magnetization vectors of the
magnetic sublattices as: $\bm{M}_F =(\bm{M}_1 + \bm{M}_2)/2$ and $\bm{M}_S =(\bm{M}_1
-\bm{M}_2)/2$, respectively, see Fig.~\ref{MsMf}.

The free energy of the CAF-ordered  $\pi$-electron spin subsystem with the sublattice magnetizations
outlined in Fig.~\ref{MsMf}, in the presence of the magnetic field is composed of the Zeeman
energy

\begin{equation}
E_Z=-\left(\bm{M}_1 + \bm{M}_2\right)\cdot \bm{H} = -2\bm{M}_F \cdot \bm{H},\label{Ez}
\end{equation}

\noindent the isotropic exchange energy

\begin{equation}
E_i=2A\left(\bm{M}_1 \cdot \bm{M}_2\right) = 2A\left[\left(\bm{M}_F\right)^2 -
\left(\bm{M}_S\right)^2\right],\label{Ei}
\end{equation}

\noindent the anisotropic exchange energy

\begin{equation}
E_a=2K_a\left(\bm{M}_1 \cdot \bm{k}\right)\left(\bm{M}_2 \cdot \bm{k}\right) =
2K_a\left[\left(\bm{M}_F\cdot\bm{k}\right)^2 -\left(\bm{M}_S\cdot\bm{k}\right)^2\right],\label{Ea}
\end{equation}

\noindent and the DM term
\begin{equation}
E_{DM}=\bm{D}\cdot(\bm{M}_1 \times \bm{M}) = 2\bm{D}\left(\bm{M}_F \times
\bm{M}_S\right),\label{EDM}
\end{equation}

\noindent where $A$ and $K_a$ are, respectively, the isotropic and anisotropic exchange constants,
$\bm{k}$ the unit vector along the anisotropic exchange easy axis, and $\bm{D}$ the DM vector.
$E_z$ is minimized when $\bm{M}_F \parallel \bm{H}$, and $E_i$ when $\bm{M}_1 = \bm{M}_2$, i.e.
when
 $\left|\bm{M}_F\right|=0$, $\left|\bm{M}_S\right|=M$ ($M$ is the magnitude of the electron spin moment in
both sublattices): the spins minimize $E_i$ by aligning in an antiparallel orientation. $E_a$ is
minimum when $\bm{M}_S\parallel \bm{k}$ because $\left| E_a\right| \ll\left| E_i\right|$, hence
$\bm{M}_F\ll \bm{M}_S$ \cite{SmPRB03, WePB93}, and the effect of $E_{DM}$ is to arrange $\bm{M}_F$
and $\bm{M}_S$ perpendicular to $\bm{D}$. The ultimate spin orientation is determined by a tradeoff
between the four contributions to the total free energy.

The crystallographic $ac$-plane is the mirror plane in the structure of  \kappamn. Symmetry
considerations, thus, require $\bm{k}$ and $\bm{D}$ vectors to lie in the $ac$-plane and $\bm{M}_F$
along the $b$-axis. Recent calculations \cite{WiJACS15} have shown that the preferable orientation
of vector $\bm{D}$ is the long axis of the BETS molecule, which is in our case directed at $\simeq
21^{\circ}$ from $a^{\ast}$ in the $ac$-plane. The exact direction of $\bm{k}$ is currently
unknown. The overall easy axis $\bm{k}_S$ of the CAF-ordered $\pi$-spin subsystem, is the
compromise between the normal to vector $\bm{D}$ and the $\bm{k}$ direction.

\begin{figure}[h!]
\includegraphics[width=0.5\linewidth]{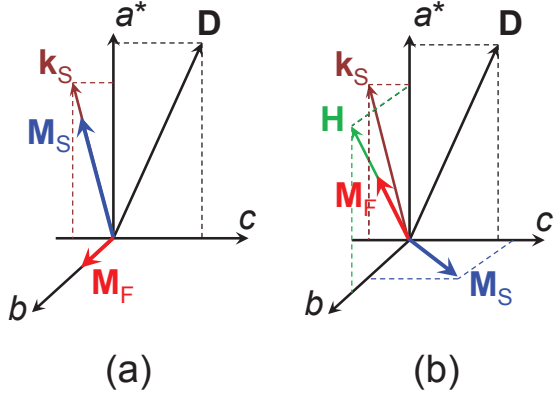}
\caption{(Color online) Arrangement of the AF sublattice moments: (a) at zero field and (b) above the
critical field of the SR transition $H_{\mathrm{kink}}$.\label{SpinArr}}
\end{figure}

Based on these considerations, one can propose a scheme of the SR transition responsible for the
kink feature in the field-dependent torque. At zero field the AF sublattice moments are arranged as
follows: $\bm{M}_F$ is along the $b$-axis and $\bm{M}_S\parallel \bm{k}_S$ is in the $ac$-plane at
some angle from $\bm{D}$, as shown in Fig.~\ref{SpinArr}(a).

As the magnetic field is applied with a strong enough component along $\bm{M}_S$, so that $\left|
E_z \right| > \left| E_a \right|$, the orientation of $\bm{M}_F$ along the $b$-axis becomes
unfavorable and it switches to (or maybe close to) the direction of the external field, producing
an abrupt change in the magnetisation anisotropy, hence a kink in the torque signal. In turn,
$\bm{M}_S$ switches to the direction perpendicular to both $\bm{M}_F$ and $\bm{D}$, see
Fig.~\ref{SpinArr}(b). Obviously, the minimum in $H_{\mathrm{kink}}$ should correspond to the
external field direction along $\bm{k}_S$. The experimental data on the angular dependence
$H_{\mathrm{kink}}(\theta)$ for different rotation planes shown in Fig.~\ref{Hkink} give a key to
understanding the orientation of $\bm{k}_S$ in the $ac$-plane. One can notice that for the
rotations around the $c$-axis and $[0\bar{1}1]$ direction, $H_{\mathrm{kink}}$ is symmetric around
$\theta=0^{\circ}$, while for the rotation around the direction perpendicular to $[0\bar{1}1]$,
which corresponds to the rotation plane closest to the $ac$-plane, the minimum in
$H_{\mathrm{kink}}(\theta)$ is shifted by $\theta \approx 5^{\circ}$  from the $a^{\ast}$
direction. For this rotation plane the projection of the field, applied at polar angle
$\theta=5^{\circ}$, on the $ac$-plane makes an angle of $\approx 4^{\circ}$ with $a^{\ast}$.
Therefore, it is likely that $\bm{k}_S$ is at some small angle from $a^{\ast}$ in the $(a, -c)$
quadrant, as shown schematically in Fig.~\ref{SpinArr}.

The suggested model of the AF spin arrangement explains the existence of the kinks in
the field dependence of the measured torque, but does not explain why the kinks are
only observed when the external field has a non-zero $b$-axis component.
For example, no kink is found for the fields exactly perpendicular to the layers, $\theta = 0^{\circ}$.
One might doubt the existence of the SR transition at this field orientation.
However, recent $^{13}$C NMR experiments confirm that
it does exist \cite{VyNHMFL}: in these experiments performed on a $^{13}$C-enriched
crystal, the mentioned SR transition at $T<T_\mathrm {MI}$ is seen as a
dramatic change in the spectrum shape right at the same values and orientations of the magnetic
field at which the kink in the field-dependent torque is observed, but also at $H\parallel a^{\ast}$
at $H\simeq7$\,T.

The apparent controversy can be resolved by taking into account that $ac$ is the mirror plane of
the crystal structure. Indeed, in this case the alignment of $\bm{M}_F$ along the $b$ and $-b$
directions is equally favorable in the absence of external field. Therefore one can expect a domain
structure to be formed with equal number of the ferromagnetic moments $\bm{M}_F$ pointing to the
directions $b$ and $-b$, respectively. When an external field exceeding the critical value is
applied exactly along the $a^{\ast}$ direction ($\theta=0^{\circ}$), the SR transition does occur,
but the change in the torque caused by switching of $\bm{M}_F$ from the $b$ direction to the
external field direction is compensated by the same process in the domains where the zero-field
moment $\bm{M}_F$ is pointing along $-b$. As a result no significant change in the total torque
happens at such field orientation. By contrast, a non-zero $b$-component of the applied field lifts
this degeneracy, and the SR transition leads to a sizeable step in the total torque.

\subsection{Interaction between $\pi$- and $d$- spin subsystems.}

So far we considered the torque features caused by the $d$- and $\pi$-spin subsystems
individually. In fact, the possibility to distinguish the contributions to the torque from the two
subsystems indicates the weakness of $\pi$--$d$ interactions, unlike, for example in $\lambda$-(BETS)$_2$FeCl$_4$,
where both $\pi$- and $d$-electron spins are antiferromagnetically ordered \cite{KonoPB05,ToPh05}
and their individual contributions to the torque can hardly be separated.

In \kappamn, the  $\pi$--$d$ interaction between the essentially paramagnetic Mn$^{2+}$
$d$-electron spin subsystem and the AF $\pi$-electron spin subsystem is apparently manifested in
the nonmonotonic behavior of the torque in the intermediate field range, below $\simeq 7.5$\,T for the
directions of the field close to the magnetization principal axes (Fig.~\ref{HDeps}), at which
the high-field/high-temperature torque is zero.

An isolated Mn$^{2+}$ spin subsystem would produce a zero torque once the field is along any
principal axis of the magnetization, since in that case the magnetization vector coincides with the
field direction. However, at temperatures below $T_\mathrm {MI}$ $\pi$-electron spins form a
long-range CAF order. Due to a finite ferromagnetic component, $M_F$, of the ordered $\pi$-electron
moments, the $d$-electron spins experience a local exchange field caused by the $\pi$--$d$
interaction. This gives rise to their nonzero magnetization even in the absence of the external
field $\bm{H}_{\pi d}$. The orientation of the zero-field magnetization of Mn$^{2+}$ depends on
details of the $\pi$--$d$ coupling and  does not need to coincide with the directions of the
magnetization principal axes. Therefore, in a small external field $\bm{H}$, even if it is applied
along a principal axis, the magnetization of the Mn$^{2+}$ subsystem is determined by the effective
field $\bm{H}_{\mathrm{eff}} = \bm{H} + \bm{H}_{\pi d}$, giving rise to a finite torque. As the
external field (along the principal axis) increases, the Zeeman energy gradually overcomes the
contribution from the $\pi$--$d$ exchange, the magnetization vector turns towards the direction of
$\bm{H}$, and the torque signal approaches zero. In our experiment this happens at $\simeq 7.5$\,T,
as one can see from Fig.~\ref{HDeps}. Thus, the observed nonmonotonic torque behavior can be
understood as a result of the $\pi$--$d$ exchange in \kappamn. Yet other manifestations of the
interaction between the two spin subsystems in this material are the violation of the Curie-Weiss
behavior of the bulk magnetization \cite{KJACS08,VyPRB11} and a sharp increase of $^1$H NMR
linewidth \cite{VyPRB11, VyJETP11} observed at $T<T_\mathrm {MI}$.

The fact that below $T_\mathrm {MI}$ the AF-ordered $\pi$-spin subsystem does not
induce the AF order in the $d$-electron Mn$^{2+}$ spin subsystem has two origins. First is the weakness
of $\pi$--$d$ coupling. While the exact value of the exchange energy is unknown as yet,
the absence of beats in Shubnikov-de Haas effect in the interval 11 to 29\,T \cite{kart17a}
sets the upper limit for it as $\lesssim 0.25$\,meV, that is $\sim 6$ times lower than
in $\lambda$-(BETS)$_2$FeCl$_4$, where both subsystems order antiferromagnetically.
The second factor suppressing the long-range order in the $d$-electron subsystem
is the polymer-type triangular structure of the Mn$^{2+}$ lattice in the anionic layers. The dicyanamide
bridges connecting Mn$^{2+}$ ions favor a direct exchange interaction within the anion layers
\cite{KJACS08}, which is likely to prevail the $\pi$--$d$ coupling, while the triangular arrangement of
Mn$^{2+}$ ions frustrates their AF-type ordering.

\section{Summary}

The anomalies found in the low-temperature magnetic torque in  \kappamn
can be understood in terms of two spatially separated and weakly interacting spin subsystems.
One subsystem is associated with $d$-electrons of the Mn$^{2+}$ ions residing in the insulating
anion layers, and the other with itinerant $\pi$-electrons in the conducting molecular layers,
which form a long-range AF structure at the Mott-insulating transition.
From the angular dependence of the high-field torque we were able to determine the directions of the
principal axes of magnetization for the Mn$^{2+}$ spin subsystem.
The sharp kink feature observed in the
field dependence of the torque in a certain angular range is interpreted as a manifestation of the
spin-reorientation transition in the $\pi$-electron subsystem. Based on the dependence of the
kink on the field orientation, a qualitative model of the canted AF spin arrangement in this subsystem
below and above the spin-orientation transition has been proposed.
Finally, the weak exchange interaction between the two subsystems is manifested in the smooth nonmonotonic
behavior of the torque at the field directions near the principal magnetization axes of Mn$^{2+}$.

\section*{Acknowledgements}

The authors gratefully acknowledge fruitful discussions with V.\,Ryazanov and S.\,Winter. The work
was supported by the German Research Foundation grant  KA~1652/4-1 and by the Russian Foundation
for Basic Research, project No.~13-02-00350.

\appendix* \section{The torque caused by the sample geometry.}

Consider an isotropic paramagnet in a shape of a general ellipsoid with semi-axes $l_a$, $l_b$ and
$l_c$, in the external field $\bm{H}_{\mathrm{e}}$,
\begin{equation}
\bm{H}_{\mathrm{e}}=H_{\mathrm{e}}[\cos\theta, \sin\theta\sin\phi, \sin\theta\cos\phi
]\,,\label{Bex}
\end{equation}
where the polar angle $\theta$ and the azimuth angle $\phi$  are reckoned from $l_a$ and $l_b$
directions, respectively. Once the material is assumed isotropic, the magnetization vector is
parallel to $\bm{H}_{\mathrm{e}}$,
\begin{equation}
\bm{M}=M[\cos\theta, \sin\theta\sin\phi, \sin\theta\cos\phi  ]\,,\label{Mdem}
\end{equation}
and saturates to a constant value at high fields.
The demagnetizing field is:
\begin{equation}
\bm{H}_{\mathrm{d}}=\mu_0\hat{n}\bm{M}\,,\label{Bd}
\end{equation}
where the demagnetizing factor
\begin{equation}
\hat{n}=\left( \begin{array}{ccc}
n_a & 0 & 0 \\
0 & n_b & 0 \\
0 & 0 & n_c \end{array} \right)\,. \label{nfac}
\end{equation}
The torque arising from the sample geometry is
\begin{eqnarray}
\bm{\tau}_{\mathrm{dem}}=V\bm{M}\times \bm{B} = V\mu_0\bm{M}\times (\bm{H}_{\mathrm{e}}-\bm{H}_{\mathrm{d}})=\nonumber \\
=2\pi \mu_0 VM^2[(n_b-n_c)\sin^2\theta \sin 2\phi ,-(n_a-n_c)\sin 2\theta\cos\phi, (n_a-n_b)\sin
2\theta\sin\phi ]\,.\label{taudem}
\end{eqnarray}

The projection of the torque on the field rotation axis, $\bm{R} = [0, -\cos\phi, \sin\phi]$, is:

\begin{equation}
\tau_{\mathrm{dem}}(\theta,\phi)= \bm{\tau}_{\mathrm{dem}}\cdot\bm{R}=2\pi\mu_0 VM^2[(n_a-n_c)\sin
2\theta\cos^2\phi+ (n_a-n_b)\sin 2\theta\sin^2\phi ]\,. \label{taudemphi}
\end{equation}

As mentioned in Sec. II, the sample dimensions are 0.08, 0.7 and 0.3\,mm along $a^{\ast}$, $b$ and
$c$ crystallographic directions, respectively. Taking these values as the ellipsoid semi-axes, and
using the approach of Refs.\,\citenum{OsPR45,BelPM06} one obtains the demagnetizing factors
$n_a = 0.755$, $n_b = 0.057$ and $n_c = 0.188$. For $M$ one can use the maximum value $46.7\times 10^3$\,A/m of
the saturated paramagnet with $L=0$, $S=5/2$, which seems to be a reasonable estimation according
to the dc magnetometry data \cite{VyPRB11}. Then for the rotation axes along $[0\bar{1}0]$
($\phi=0$), [001] ($\phi=90^{\circ}$), $[0\bar{1}1]$($\phi=55^{\circ}$), and the perpendicular to
$[0\bar{1}1]$ ($\phi=145^{\circ}$) Eq.~\ref{taudemphi} gives (in units $10^{-7}$\,N$\cdot$m)

\begin{subequations}
\label{taudemrot}

\begin{equation}
\tau_{\mathrm{dem}}(\phi=0^{\circ})=0.13\sin 2\theta, \label{taudem0}
\end{equation}

\begin{equation}
\tau_{\mathrm{dem}}(\phi=90^{\circ})=0.16\sin 2\theta, \label{taudem90}
\end{equation}

\begin{equation}
\tau_{\mathrm{dem}}(\phi=55^{\circ})=0.15\sin 2\theta, \label{taudem55}
\end{equation}

\begin{equation}
\tau_{\mathrm{dem}}(\phi=145^{\circ})=0.14\sin 2\theta. \label{taudem145}
\end{equation}

\end{subequations}

\end{document}